\documentclass[a4paper,12pt]{article}

\usepackage[utf8]{inputenc}

\usepackage[margin=1in,includefoot,footskip=30pt]{geometry}

\usepackage{amsmath}
\usepackage{amssymb}
\usepackage{hyperref}
\usepackage{graphicx}
\usepackage{siunitx}

\numberwithin{equation}{section}

\DeclareMathOperator{\tr}{tr}
\DeclareMathOperator{\sgn}{sgn}
\newcommand{\GRACE}{\textsc{Grace}}

\usepackage{color}
\providecommand{\color}[1]{}

\begin{document}

\title{Higgs triplet extension of GRACE}
\author{
  Yusaku Kouda,\!$^1$\thanks{s91661@cc.seikei.ac.jp} \
  Tadashi Kon,\!$^1$\thanks{kon@st.seikei.ac.jp} \
  Yoshimasa Kurihara$^2$\thanks{kurihara@post.kek.jp} \\
  and Takahiro Ueda$^1$\thanks{tueda@st.seikei.ac.jp}
  \bigskip \\
  \itshape $^1$Faculty of Science and Technology, Seikei University, \\
  \itshape 3-3-1 Kichijoji-Kitamachi, Musashino, Tokyo 180-8633, Japan
  \bigskip \\
  \itshape $^2$High Energy Accelerator Organization (KEK), \\
  \itshape Tsukuba, Ibaraki 305-0801, Japan
}
\date{}
\maketitle

\abstract{
  Much theoretical effort and automatization are required
  to confront new physics models with experimental data
  for many types of particle reactions at future colliders.
  In this context,
  we extend \GRACE{}, an automatic calculation system for invariant amplitudes,
  to incorporate particles and interactions in the Georgi--Machacek model.
  With the extended \GRACE{} system, we study fermiophobic Higgs boson
  production processes at $e^+ e^-$ and $e^- e^-$ colliders in the model.
  The results show some advantages of $e^- e^-$ colliders over
  $e^+ e^-$ colliders for new physics search and thus its complementary role.
}

\section{Introduction}

The discovery of a Higgs boson in 2012~\cite{ATLAS:2012yve,CMS:2012qbp}
completed the last piece missing from
the Standard Model (SM) of particle physics.
Nevertheless, there is still much that cannot be explained within the SM\@.
Some of the open questions are addressed by extending the Higgs sector
or by models that force us to extend that sector
(see Refs.~\cite{Ivanov:2017dad,Dawson:2018dcd} for
recent reviews on models with extended Higgs sectors).
Vast possibilities for non-minimal Higgs sectors will be tested by
comparing future experimental data with model predictions.
To obtain a quantitative prediction for the collider phenomenology,
one may use tools dedicated to specific models (e.g.,~\cite{Hartling:2014xma,Kanemura:2019slf})
or general-purpose tools written in a model-independent way
(see Ref.~\cite{Fontes:2019wqh} and references therein).

The aim of this paper is twofold.
Firstly, we discuss extending the public version of
the \GRACE{} system~\cite{Yuasa:1999rg,Fujimoto:2002sj,Grace:2006}
for models with exotic Higgs particles, beyond isospin doublets.
Here, we focus on the Georgi--Machacek (GM) model~\cite{Georgi:1985nv} as a concrete example
that contains isospin triplets while keeping the custodial symmetry naturally~\cite{Chanowitz:1985ug}
and has rich collider phenomenology, e.g.,~%
\cite{%
  Gunion:1989ci,%
  Chiang:2012cn,%
  Chiang:2015rva%
}.
In fact, one can implement a new model with \GRACE{} by preparing a set of files describing the model,
i.e., particle contents, interaction vertices, model parameters, etc.
Once the model files are established,
then in principle one can compute any tree-level decay and scattering amplitudes.

Secondly, using \GRACE{} equipped with the GM model,
we compute production cross sections of fermiophobic custodial 5-plet Higgs bosons
($H_5^{\pm\pm}$, $H_5^\pm$, $H_5^0$) at future $e^+ e^-$ and $e^- e^-$ colliders.
It is known that a clear signature of physics beyond the SM at $e^- e^-$ colliders
may come from $W^- W^-$ production~\cite{Rizzo:1982kn,Barger:1994wa},
which is the counterpart of the same-sign $W^\pm W^\pm$ production measurement
at the LHC~\cite{CMS:2017fhs,ATLAS:2019cbr}.
Notably, Ref.~\cite{Barger:1994wa} considered the resonant effect of $H_5^{--}$
on $W^- W^-$ production cross section
at $e^- e^-$ colliders in the GM model.
See Ref.~\cite{Feng:1999zv} for other physics cases at $e^- e^-$ colliders.
Comparing the production cross sections in $e^+ e^-$ collisions~\cite{Gunion:1989ci,Chiang:2015rva}
with those in $e^- e^-$ collisions, we will see
advantages of the latter for the 5-plet Higgs boson search.

The GM model adds isospin triplets in the scalar sector,
the vacuum expectation values (VEVs) of which could give raise to Majorana neutrino masses
by the type-II seesaw mechanism%
~\cite{Cheng:1980qt,Schechter:1980gr,Magg:1980ut,Lazarides:1980nt,Mohapatra:1980yp}
as other Higgs triplet models,
while naturally maintaining the electroweak $\rho$ parameter as unity at the tree level.%
\footnote{%
  See, however, Ref.~\cite{Gunion:1990dt} for $\rho$ at the one-loop level.
}
Higgs bosons in the same custodial multiplet have degenerated masses at the tree level~%
\cite{%
  Chanowitz:1985ug,
  Chiang:2012cn
},
which may make the model easily distinguishable from other models.
The SM-like Higgs boson's couplings to the weak gauge bosons can be larger than
those in the SM~\cite{Logan:2010en},
which is favored by the slightly larger central values measured
by the ATLAS experiment~\cite{ATLAS:2019nkf} (but see also the CMS result~\cite{CMS:2018uag}).
The model allows the strong first-order electroweak phase transition%
~\cite{Chiang:2014hia,Zhou:2018zli},
thus provides a scenario of electroweak baryogenesis.
Theoretical constraints on the parameter space in the GM model are examined in
Refs.~\cite{Aoki:2007ah,Hartling:2014zca,Krauss:2017xpj,Krauss:2018orw}.
See Refs.~\cite{Li:2017daq,Chiang:2018cgb,Ghosh:2019qie,Ismail:2020zoz,CMS:2021wlt}
for recent experimental constraints and parameter space analyses.

This paper is organized as follows.
In Sec.~\ref{sec:model}, we give the definition of the GM model
to clarify our conventions.
In Sec.~\ref{sec:grace}, we discuss an extension of the \GRACE{} system to the GM model.
Numerical results for custodial 5-plet Higgs boson production cross sections
are shown in Sec.~\ref{sec:result}, which exhibit some advantages of $e^- e^-$ colliders over
$e^+ e^-$ colliders for new physics searches.
Sec.~\ref{sec:summary} is devoted to the summary.

\section{Model}
\label{sec:model}

Here, we briefly recapitulate the GM model~\cite{Georgi:1985nv}
with the focus on the physical Higgs bosons.%
\footnote{%
  Our conventions are almost the same as those in Ref.~\cite{Chiang:2012cn}
  except that we do not introduce $M_1^2$ and $M_2^2$ in Eq.~(2.9) of Ref.~\cite{Chiang:2012cn}
  and we use $(\chi^{++})^*$, $(\xi^+)^*$, $(\phi^+)^*$, etc.\ in this section
  rather than $\chi^{--}$, $\xi^-$, $\phi^-$, etc.
  Note that there are many conventions used in the literature for the parameterization of
  the Higgs potential, (signs of) mixing angles and so on.
}
The Higgs sector of the GM model contains one real isospin triplet field $\xi$
with hypercharge $Y=0$ and one complex isospin triplet field $\chi$ with $Y=1$,
as well as the usual complex isospin doublet field $\phi$ with $Y=1/2$ in the SM\@.
It is convenient to express these scalar fields as
$SU(2)_L \times SU(2)_R$ bi-doublet and bi-triplet:
\begin{equation}
  \Phi
  =
  \begin{pmatrix}
    \phantom{-} (\phi^0)^* & \phi^+ \\
    - (\phi^+)^* & \phi^0
  \end{pmatrix} ,
  \qquad
  \Delta = \begin{pmatrix}
    \phantom{-} (\chi^0)^* & \phantom{-} \xi^+ & \chi^{++} \\
    -(\chi^+)^* & \phantom{-} \xi^0 & \chi^+ \\
    (\chi^{++})^* & -(\xi^+)^* & \chi^0
  \end{pmatrix} .
\end{equation}

The kinetic terms for the scalar fields are given as
\begin{equation}
  \mathcal{L}_\text{kin}
  = \frac{1}{2} \tr\left[ (\mathcal{D}_\mu \Phi)^\dagger (\mathcal{D}^\mu \Phi) \right]
  + \frac{1}{2} \tr\left[ (\mathcal{D}_\mu \Delta)^\dagger (\mathcal{D}^\mu \Delta) \right] ,
\end{equation}
where the covariant derivatives are defined by
\begin{align}
  \mathcal{D}_\mu \Phi &= \partial_\mu \Phi + i g \frac{\tau^a}{2} W^a_\mu \Phi - i g' B_\mu \Phi \frac{\tau^3}{2} , \\
  \mathcal{D}_\mu \Delta &= \partial_\mu \Delta + i g t^a W^a_\mu \Delta - i g' B_\mu \Delta t^3 ,
\end{align}
with $\tau^a$ being the Pauli matrices and $t^a$ being generators for $SU(2)$ triplet representations,
which one can choose
\begin{equation}
  t^1 = \frac{1}{\sqrt{2}} \begin{pmatrix}
    0 & 1 & 0 \\
    1 & 0 & 1 \\
    0 & 1 & 0
  \end{pmatrix} ,
  \qquad
  t^2 = \frac{1}{\sqrt{2}} \begin{pmatrix}
    0 & -i & 0 \\
    i & 0 & -i \\
    0 & i & 0
  \end{pmatrix} ,
  \qquad
  t^3 = \begin{pmatrix}
    1 & 0 & 0 \\
    0 & 0 & 0 \\
    0 & 0 & -1
  \end{pmatrix} .
\end{equation}

The most general Higgs potential preserving the $SU(2)_L \times SU(2)_R$ symmetry%
~\cite{Aoki:2007ah} is given in our conventions as
\begin{equation}
  \begin{split}
    V_H
    ={}&
    m_1^2 \tr(\Phi^\dagger \Phi)
    + m_2^2 \tr(\Delta^\dagger \Delta)
    + \lambda_1 \left[ \tr(\Phi^\dagger \Phi) \right]^2
    + \lambda_2 \left[ \tr(\Delta^\dagger \Delta) \right]^2
    \\&
    + \lambda_3 \tr\left[ (\Delta^\dagger \Delta)^2 \right]
    + \lambda_4 \tr(\Phi^\dagger \Phi) \tr(\Delta^\dagger \Delta)
    + \lambda_5 \tr\left( \Phi^\dagger \frac{\tau^a}{2} \Phi \frac{\tau^b}{2} \right)
      \tr(\Delta^\dagger t^a \Delta t^b)
      \\&
    + \mu_1 \tr\left( \Phi^\dagger \frac{\tau^a}{2} \Phi \frac{\tau^b}{2} \right)
      (P^\dagger \Delta P)^{ab}
    + \mu_2 \tr(\Delta^\dagger t^a \Delta t^b)
      (P^\dagger \Delta P)^{ab} ,
  \end{split}
\end{equation}
where the matrix $P$ is defined as
\begin{equation}
  P = \begin{pmatrix}
    - \frac{1}{\sqrt{2}} & \frac{i}{\sqrt{2}} & 0 \\
    0 & 0 & 1 \\
    \frac{1}{\sqrt{2}} & \frac{i}{\sqrt{2}} & 0
  \end{pmatrix} .
\end{equation}
Note that if one sets both the trilinear couplings $\mu_1$ and $\mu_2$ to zero,
which leads to the conventional $\mathbb{Z}_2$ symmetric version of the GM model~\cite{Chanowitz:1985ug},
then the model is already highly constrained~\cite{Das:2018vkv}.

The spontaneous symmetry breaking is triggered by the VEVs
of the neutral components of the scalar fields:
\begin{equation}
  \langle \Phi \rangle
  =
  \frac{v_\phi}{\sqrt{2}}
  \begin{pmatrix}
    1 & 0 \\
    0 & 1
  \end{pmatrix} ,
  \qquad
  \langle \Delta \rangle
  =
  v_\Delta
  \begin{pmatrix}
    1 & 0 & 0 \\
    0 & 1 & 0 \\
    0 & 0 & 1
  \end{pmatrix} .
\end{equation}
The weak bosons acquire their masses as
\begin{equation}
  m_W^2 = m_Z^2 \cos^2\theta_W = \frac{g^2}{4} v^2 ,
\end{equation}
where $v^2 = 1 / (\sqrt{2} G_F) \approx (\SI{246}{GeV})^2$ consists of contributions from the two VEVs:
\begin{equation}
  v^2 = v_\phi^2 + 8 v_\Delta^2 .
\end{equation}
We define the doublet-triplet mixing angle $\theta_H$ as
$\tan\theta_H = 2\sqrt{2} v_\Delta /v_\phi$,
and use abbreviations $s_H = \sin\theta_H$, $c_H = \cos\theta_H$ and $t_H = \tan\theta_H$.
The minimization conditions of the Higgs potential relate the VEVs
to the other dimensionful parameters in the potential, which allows one to
eliminate $m_1^2$ and $m_2^2$.
Note that the VEVs of the neutral components of the triplet fields
are taken to be the same value, $\langle \xi^0 \rangle = \langle \chi^0 \rangle = v_\Delta$.
This preserves the diagonal custodial $SU(2)_V$ symmetry after its spontaneous breaking
$SU(2)_L \times SU(2)_R \to SU(2)_V$
and keeps the electroweak $\rho$ parameter, $\rho \equiv m_W^2 / (m_Z^2 \cos^2\theta_W)$, as unity at the tree level.


Given the exact form of the Higgs potential and the symmetry breaking pattern,
one can determine the relations between the weak eigenstates and the mass eigenstates of the scalar fields.
The preserved custodial symmetry classifies the states as $SU(2)_V$ multiplets;
the field $\Phi$ contains a 3-plet and a singlet
while the field $\Delta$ contains a 5-plet, a 3-plet and a singlet.
In general, the two 3-plets coming from $\Phi$ and $\Delta$ mix each other.
The same happens for the two singlets.
Consequently, the components fields in $\Phi$ and $\Delta$ are related to
physical Higgs boson states $(H_5^{++}, H_5^+, H_5^0)$, $(H_3^+, H_3^0)$, $H_1^0$, $h$
and Nambu--Goldstone boson states $(G^+, G^0)$ as follows:
\begin{align}
  \chi^{++} &= H_5^{++} ,
  \label{eq:doubly-charged-higgs}
  \\
  \begin{pmatrix}
    \phi^+ \\ \xi^+ \\ \chi^+
  \end{pmatrix}
  &=
  \begin{pmatrix}
    1 & 0 & \phantom{-} 0 \\
    0 & \frac{1}{\sqrt{2}} & - \frac{1}{\sqrt{2}} \\
    0 & \frac{1}{\sqrt{2}} & \phantom{-} \frac{1}{\sqrt{2}}
  \end{pmatrix}
  \begin{pmatrix}
    c_H & -s_H & 0\\
    s_H & \phantom{-} c_H & 0 \\
    0 & \phantom{-} 0 & 1
  \end{pmatrix}
  \begin{pmatrix}
    G^+ \\ H_3^+ \\ H_5^+
  \end{pmatrix} ,
  \label{eq:singly-charged-higgs}
  \\
  \begin{pmatrix}
    \phi_i \\ \chi_i
  \end{pmatrix}
  &=
  \begin{pmatrix}
    c_H & -s_H \\
    s_H & \phantom{-} c_H
  \end{pmatrix}
  \begin{pmatrix}
    G^0 \\ H_3^0
  \end{pmatrix} ,
  \\
  \begin{pmatrix}
    \phi_r \\ \xi_r \\ \chi_r
  \end{pmatrix}
  &=
  \begin{pmatrix}
    1 & 0 & \phantom{-} 0 \\
    0 & \frac{1}{\sqrt{3}} & - \sqrt{\frac{2}{3}}  \\
    0 & \sqrt{\frac{2}{3}} & \phantom{-} \frac{1}{\sqrt{3}}
  \end{pmatrix}
  \begin{pmatrix}
    c_\alpha & -s_\alpha & 0\\
    s_\alpha & \phantom{-} c_\alpha & 0 \\
    0 & \phantom{-} 0 & 1
  \end{pmatrix}
  \begin{pmatrix}
    h \\ H_1^0 \\ H_5^0
  \end{pmatrix} .
  \label{eq:cp-even-higgs}
\end{align}
Here, we have parameterized the neutral component fields $\phi^0$, $\chi^0$ and $\xi_0$ as
\begin{equation}
  \phi^0 = \frac{1}{\sqrt{2}} (\phi_r + v_\phi + i \phi_i) , \qquad
  \chi^0 = \frac{1}{\sqrt{2}} (\chi_r + i \chi_i) + v_\Delta , \qquad
  \xi^0 = \xi_r + v_\Delta .
\end{equation}
Another mixing angle $\alpha$ has been introduced for the $SU(2)_V$ singlet states
(via $s_\alpha \equiv \sin\alpha$ and $c_\alpha \equiv \cos\alpha$), which diagonalizes
a submatrix
\begin{equation}
  M^2_{\begin{subarray}{l} \text{CP-even} \\ \text{singlet} \end{subarray}} =
  \begin{pmatrix}
    M^2_{11} & M^2_{12} \\
    M^2_{12} & M^2_{22}
  \end{pmatrix} ,
\end{equation}
where
\begin{align}
  M^2_{11} &= 8 c_H^2 \lambda_1 v^2 , \\
  M^2_{22} &= s_H^2 (3 \lambda_2 + \lambda_3) v^2
    - \frac{1}{\sqrt{2}} \frac{c_H^2}{s_H} \mu_1 v
    + \frac{3}{\sqrt{2}} s_H \mu_2 v , \\
  M^2_{12} &= \sqrt{\frac{3}{2}} s_H c_H (2 \lambda_4 + \lambda_5) v^2
    + \frac{\sqrt{3}}{2} c_H \mu_1 v ,
\end{align}
and is determined by
\begin{align}
  \tan2\alpha &= \frac{2 M^2_{12}}{M^2_{11} - M^2_{22}}.
\end{align}
Without loss of generality,
one can choose the state $h$ in such a way that $h$ is lighter than $H_1^0$
by fixing the sign of $\alpha$ as
\begin{align}
  \sgn(\sin2\alpha) &= - \sgn\left(M^2_{12} \right),
\end{align}
see the mass eigenvalues below.
In this paper, we identify the lighter singlet state $h$ as the \SI{125}{GeV} Higgs boson.

The masses of the physical Higgs boson states are as follows:
\begin{align}
  \begin{split}
    m_{H_5}^2
    &\equiv m_{H_5^{++}}^2 = m_{H_5^+}^2 = m_{H_5^0}^2 \\
    &= \left( s_H^2 \lambda_3 - \frac{3}{2} c_H^2 \lambda_5 \right) v^2
      - \frac{1}{\sqrt{2}} \frac{c_H^2}{s_H} \mu_1 v
      - 3 \sqrt{2} s_H \mu_2 v ,
  \end{split} \\
  \begin{split}
    m_{H_3}^2
    &\equiv m_{H_3^+}^2 = m_{H_3^0}^2 \\
    &= - \frac{1}{2} \lambda_5 v^2
      - \frac{1}{\sqrt{2}s_H} \mu_1 v ,
  \end{split} \\
  \begin{split}
    m_{H_1}^2
    &\equiv m_{H_1^0}^2 \\
    &= M^2_{11} s_\alpha^2 + M^2_{22} c_\alpha^2 - 2 M^2_{12} s_\alpha c_\alpha \\
    &= \frac{1}{2} \left[ M^2_{11} + M^2_{22} + \sqrt{(M^2_{11} - M^2_{22})^2 + 4 (M^2_{12})^2} \right] ,
  \end{split} \\
  \begin{split}
    m_h^2
    &= M^2_{11} c_\alpha^2 + M^2_{22} s_\alpha^2 + 2 M^2_{12} s_\alpha c_\alpha \\
    &= \frac{1}{2} \left[ M^2_{11} + M^2_{22} - \sqrt{(M^2_{11} - M^2_{22})^2 + 4 (M^2_{12})^2} \right] .
  \end{split}
\end{align}
Due to the custodial symmetry, physical Higgs bosons in the same $SU(2)_V$ multiplet
have the degenerated mass at the tree level.
It is a straightforward task to express the five dimensionless parameters
$\lambda_1, \dots, \lambda_5$ in terms of the physical Higgs boson masses
$m_{H_5}$, $m_{H_3}$, $m_{H_1}$, $m_h$ and the mixing angle $\alpha$~\cite{Chiang:2012cn}.
To implement the GM model in \GRACE{},
all the particles should be given in their mass eigenstates
and interaction vertices are written in these states with physical parameters.
We choose a set of independent physical parameters originating from the Higgs potential as follows:
the 4 physical Higgs boson masses ($m_{H_5}$, $m_{H_3}$, $m_{H_1}$, $m_h$),
the VEV $v$ and the 2 mixing angles ($\theta_H$ and $\alpha$)
and the 2 dimensionful trilinear couplings ($\mu_1$ and $\mu_2$).

The isospin doublet field $\phi = (\phi^+, \phi^0)$ has the Yukawa interactions with quarks and leptons:
\begin{equation}
  \mathcal{L}_Y
  =
  - \overline{Q_L'} Y_u \tilde{\phi} u_R'
  - \overline{Q_L'} Y_d \phi d_R'
  - \overline{L_L'} Y_e \phi e_R' + \text{h.c.},
\end{equation}
where $\tilde{\phi} = i \tau^2 \phi^*$ and the fermion fields with primes are in the weak-eigenstate basis.
Neglecting Majorana neutrino masses,
which arise from the Yukawa interactions between the complex isospin triplet $\chi$ and the lepton doublets,
one finds the Yukawa interactions in the GM model having the following form for every generations
in the mass-eigenstate basis:
\begin{equation}
  \begin{split}
    \mathcal{L}_Y
    \supset
    &
    - \sum_{f=u,d,e} \frac{m_f}{v} \left[
      \frac{c_\alpha}{c_H} \bar{f} f h
      - \frac{s_\alpha}{c_H} \bar{f} f H_1^0
      + i \text{Sign}(f) t_H \bar{f} \gamma_5 f H_3^0
    \right]
    \\
    &
    + \left\{
      - \frac{\sqrt{2} V_{ud}}{v} t_H \bar{u}
        (m_u \mathcal{P}_L - m_d \mathcal{P}_R) d H_3^+
      + \frac{\sqrt{2} m_e}{v} t_H \bar{\nu} \mathcal{P}_R e H_3^+
      + \text{h.c.}
    \right\}.
  \end{split}
\end{equation}
Here $V_{ud}$ is an appropriate element of the Cabibbo--Kobayashi--Maskawa matrix,
the projection operator is defined as $\mathcal{P}_{R,L} = (1 \pm \gamma_5) / 2$
and $\text{Sign}(f)$ is given by
\begin{equation}
  \text{Sign}(f) =
  \begin{cases}
    +1 & \text{for } f = u, \\
    -1 & \text{for } f = d, e.
  \end{cases}
\end{equation}
Note that the custodial 5-plet Higgs bosons ($H_5^{++}$, $H_5^+$, $H_5^0$
in Eqs.~\eqref{eq:doubly-charged-higgs},~\eqref{eq:singly-charged-higgs} and~\eqref{eq:cp-even-higgs})
are linear combinations of the component fields in the isospin triplets $\xi$ and $\chi$.
Therefore, they do not have the usual Yukawa interactions with fermions as in the SM and become fermiophobic.

\section{Extension of the GRACE system}
\label{sec:grace}

The \GRACE{} system is a set of programs for automatic calculation of invariant amplitudes in quantum field theory developed by the Minami-Tateya collaboration at KEK\@.
The user can obtain numerical results of various cross sections and decay widths by selecting the appropriate phase space option.
The public version of \GRACE{} can perform calculations in the SM and the minimal supersymmetric standard model~\cite{Yuasa:1999rg,Fujimoto:2002sj}.
This system can be extended by adding particles and interactions to a few text files describing the model-dependent part as appropriate.
For example, an extension to the two-Higgs-doublet model and an analysis using it was done in Ref.~\cite{Kon:2018vmv}.

We extended \GRACE{} for the SM to perform calculations in the GM model.
Specifically, we incorporated the set of Higgs particles ($H_5^{\pm\pm}$, $H_5^{\pm}$, $H_5^{0}$, $H_3^{\pm}$, $H_3^{0}$, $H_1^{0}$, $h^0$) and their interactions with gauge bosons and fermions.%
\footnote{%
  Three- and four-scalar vertices are not needed for the numerical results presented in the next section.
  We leave implementation of such interaction vertices and analysis that can be affected by them for future work.
}
We selected the following 30 types of two-body decay widths and $2\to 2$ cross sections, and systematically tested the correctness of these incorporations by checking whether the analytical results and \GRACE{} results agree by at least 10 orders of magnitude.
\begin{alignat}{3}
  & H_5^{++}  && \to W^+W^+, \quad W^+ H_3^+, \\
  & H_5^{+}   && \to W^+Z, \quad Z H_3^+, \quad W^+ H_3^0, \\
  & H_5^{0}   && \to W^+W^-, \quad Z Z, \quad Z H_3^0, \quad W^+ H_3^-, \\
  & H_3^{+}   && \to t\bar{b}, \quad c\bar{s}, \quad W^+ H_1^0, \quad W^+ h^0, \\
  & H_3^{0}   && \to t\bar{t}, \quad b\bar{b}, \quad Z H_1^0, \quad Z h^0, \\
  & H_1^{0}   && \to t\bar{t}, \quad b\bar{b}, \quad W^+W^-, \quad Z Z, \\
  & h^{0}     && \to b\bar{b}, \\
  & e^+e^-    && \to H_5^{++}H_5^{--}, \quad H_5^{+}H_5^{-}, \quad H_3^{+}H_3^{-}, \quad Z h^{0}, \\
  & e^+ \nu_e && \to H_5^{++}H_5^{-}, \quad H_5^{+}H_5^{0}, \quad H_3^{+}H_3^{0}, \quad W^+ h^{0}.
\end{alignat}
Formulae for cross sections and decay widths used in the analytical calculations, as well as expressions for the interaction coefficients, are provided in Appendix~\ref{sec:formulae}.

\begin{figure}
  \centering
  \includegraphics[width=.95\textwidth]{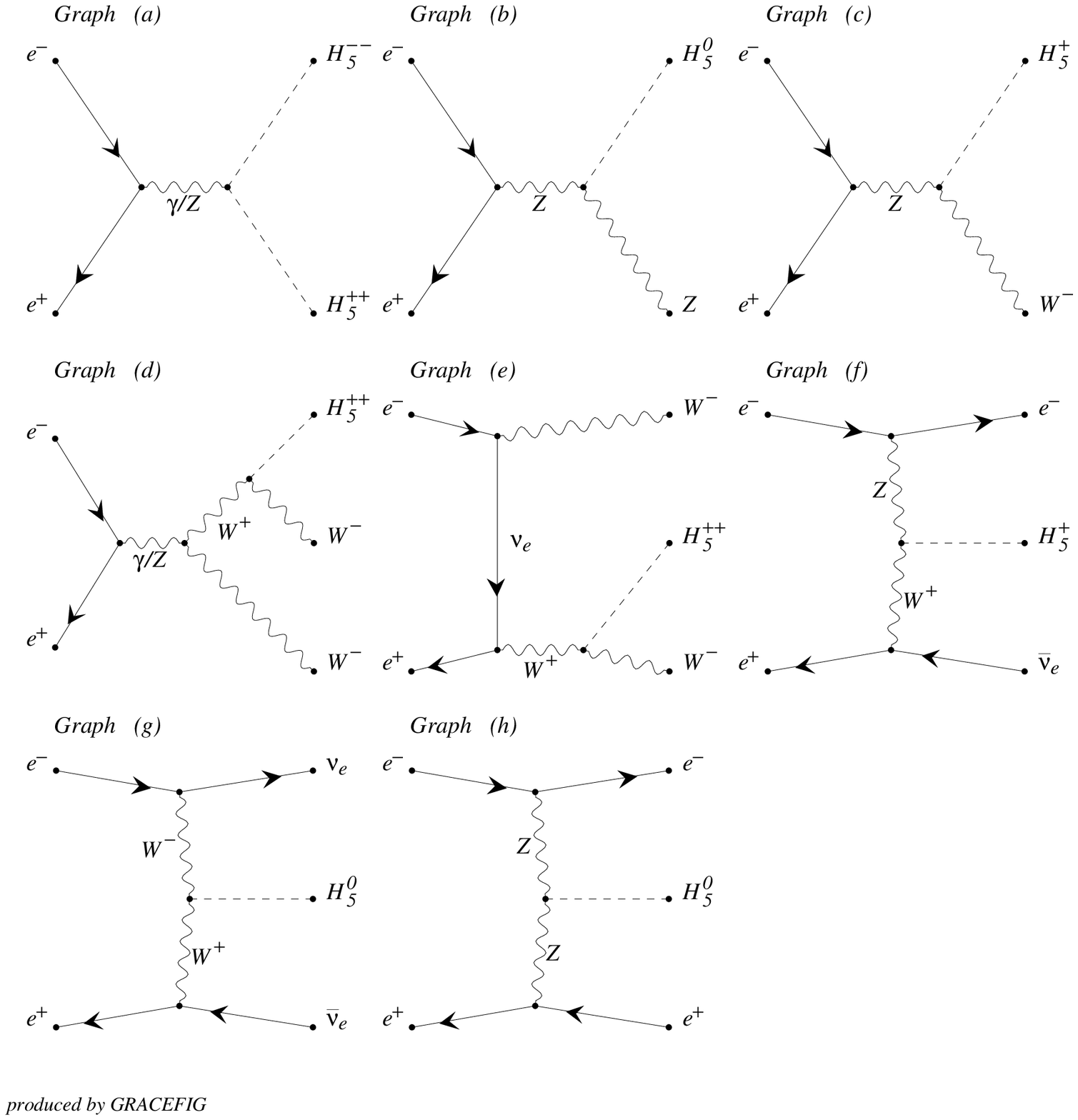}
  \caption{%
    Typical Feynman diagrams for the 5-plet Higgs boson production in $e^+ e^-$ collisions:
    (a)~$e^+ e^- \to H_5^{++} H_5^{--}$ (similar diagrams for $e^+ e^- \to H_5^+ H_5^-$),
    (b)~$e^+ e^- \to Z H_5^0$,
    (c)~$e^+ e^- \to W^- H_5^+$,
    (d)~and~(e)~$e^+ e^- \to W^- W^- H_5^{++}$,
    (f)~$e^+ e^- \to e^- \bar{\nu}_e H_5^+$,
    (g)~$e^+ e^- \to \nu_e \bar{\nu}_e H_5^0$ and
    (h)~$e^+ e^- \to e^+ e^- H_5^0$.
    The diagram~(a) is classified as pair production,
    the diagrams~(b)--(e) are VBA type, 
    and the diagrams~(f)--(h) are VBF type.
  }
  \label{fig:eE-diag}
\end{figure}

In addition, we have reproduced the production cross sections of several Higgs bosons at $e^+e^-$ colliders that are shown in Fig.~5 of Ref.~\cite{Chiang:2015rva}.
The list of concrete processes is as follows:
\begin{align}
  e^+e^- & \to H_5^{++}H_5^{--}, \quad H_5^{+}H_5^{-}, \\
  e^+e^- & \to Z H_5^0,  \quad W^- H_5^+ + \text{c.c.}, \quad W^- W^- H_5^{++} + \text{c.c.}, \\
  e^+e^- & \to e^- \bar{\nu}_e H_5^{+} + \text{c.c.}, \quad \nu_e \bar{\nu}_e H_5^{0}, \quad e^+ e^- H_5^{0}.
\end{align}
Typical Feynman diagrams contributing to these processes are shown in Fig.~\ref{fig:eE-diag}.
The first type includes pair production of the doubly- and singly-charged Higgs bosons.
The second and third types involve vector boson associated (VBA) and vector boson fusion (VBF) processes, respectively.
One can also find calculations for these production cross sections in Ref.~\cite{Gunion:1989ci}.

The \GRACE{} model files for the GM model are available from the authors upon request.

\begin{figure}
  \centering
  \includegraphics[width=.70\textwidth]{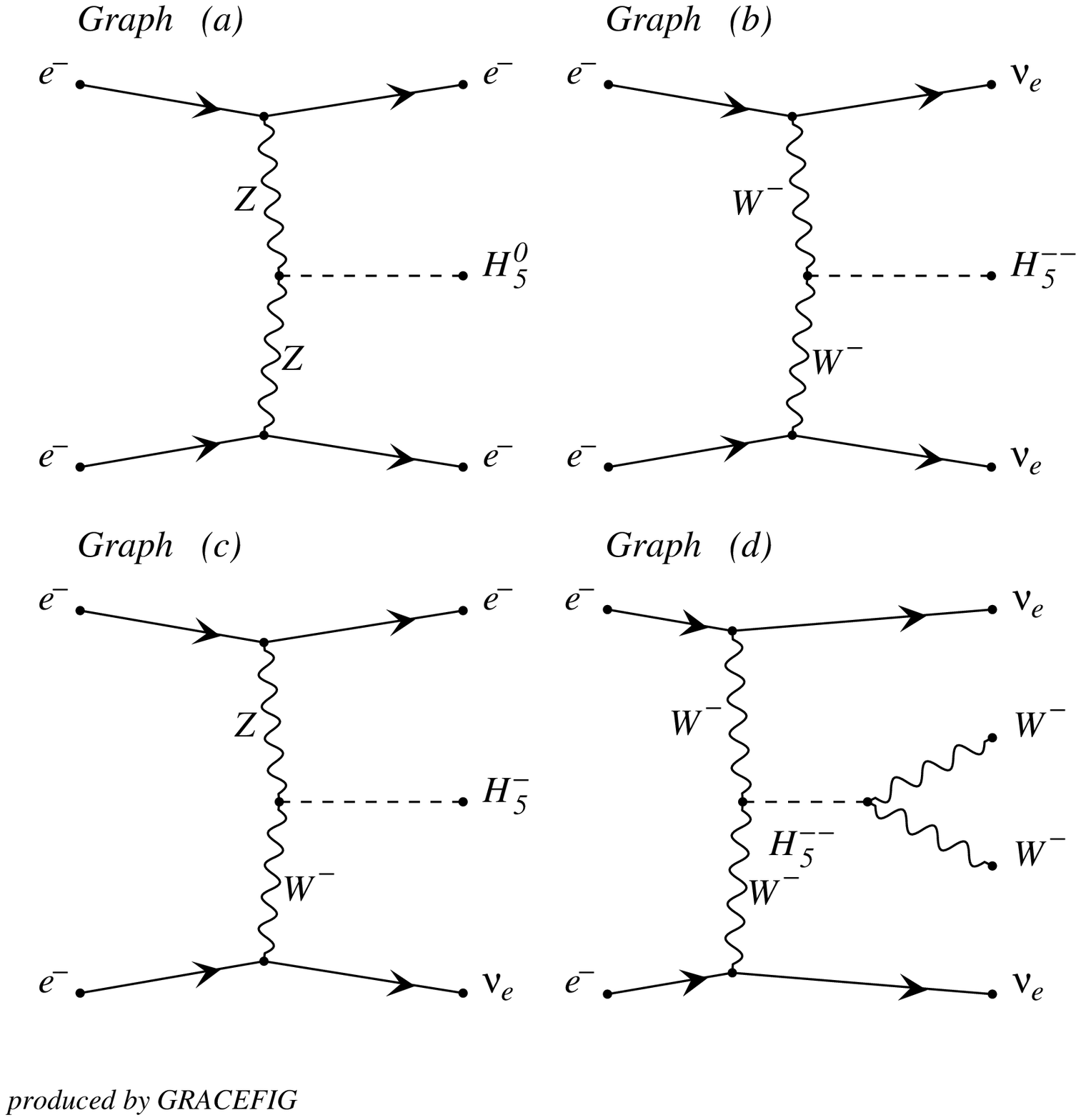}
  \caption{%
    Typical Feynman diagrams contributing to:
    (a)~$e^- e^- \to e^- e^- H_5^0$,
    (b)~$e^- e^- \to \nu_e \nu_e H_5^{--}$,
    (c)~$e^- e^- \to e \nu_e H_5^-$ and
    (d)~$e^- e^- \to \nu_e \nu_e W^- W^-$.
  }
  \label{fig:ee-diag}
\end{figure}

\section{Numerical results}
\label{sec:result}

In this section, we present production cross sections of
the custodial 5-plet Higgs bosons at future $e^+ e^-$ and $e^- e^-$ colliders,
computed by using \GRACE{} implementing the GM model.
The center-of-mass energy $\sqrt{s}$ is assumed to be \SI{0.5}{TeV}.
Typical Feynman diagrams appearing in $e^- e^-$ processes we consider are shown in Fig.~\ref{fig:ee-diag}.
Among the unknown model parameters in the GM model, only the 5-plet Higgs boson mass $m_{H_5}$
and the doublet-triplet mixing angle $\theta_H$ affects these cross sections.
The latest CMS result~\cite{CMS:2021wlt} excludes $s_H \equiv \sin\theta_H = 2 \sqrt{2} v_\Delta / v$
greater than 0.20--0.35 at the 95\% confidence level for a wide range of $m_{H_5}$.
On the other hand, some points of $m_{H_5} \lessapprox \SI{140}{GeV}$ in the full parameter space are
already excluded by Drell--Yan production of $H_5^0$ decaying to diphoton~\cite{Ismail:2020zoz}.
Therefore, as a typical benchmark point,
we take $v_\Delta = \SI{10}{GeV}$ ($s_H \approx 0.11$)
with varying $m_{H_5}$ from \SI{150}{GeV} to around $\sqrt{s}$.
In addition to the above experimental constraints,
these points satisfy the theoretical constraints in Ref.~\cite{Chiang:2012cn}.

\begin{figure}
  \centering
  \includegraphics[width=.8\textwidth]{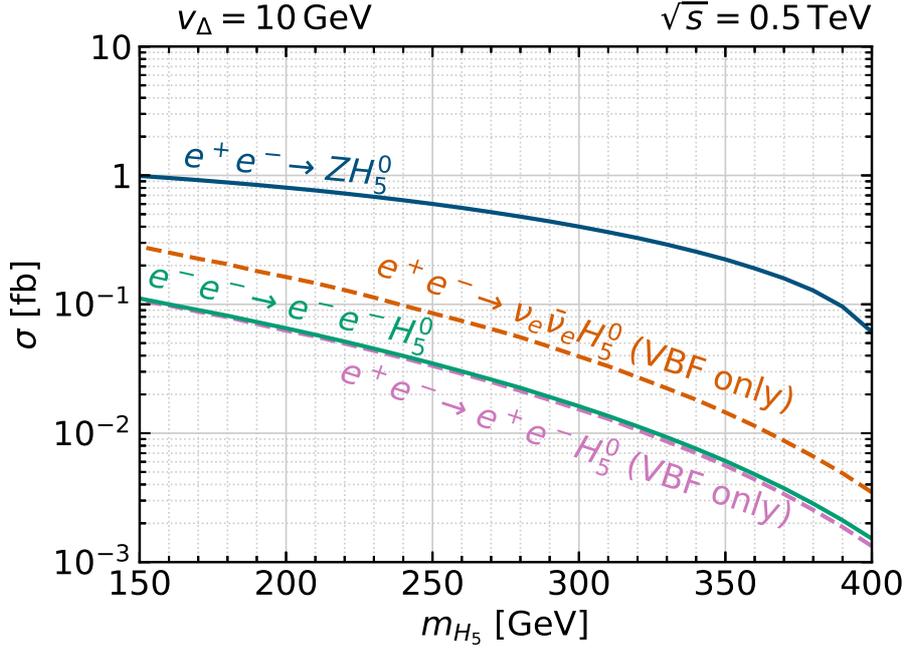}
  \caption{%
    The $H_5^0$ production cross sections at $e^+ e^-$ and $e^- e^-$ colliders.
    For the dashed curves ($e^+ e^- \to \nu_e \bar{\nu}_e H_5^0$ and $e^+ e^- \to e^+ e^- H_5^0$),
    only VBF-type diagrams are taken into account in the computations.
  }
  \label{fig:h50-prod}
\end{figure}

Fig.~\ref{fig:h50-prod} shows the mass $m_{H_5}$ dependence of the total cross section for several neutral $H_5^0$ particle production processes in $e^+e^-$ and $e^-e^-$ collisions.
Here, $e^+e^- \to Z H_5^0$ has only a VBA-type Feynman diagram, except negligible ones.
The other $e^+e^-$ processes contain Feynman diagrams of both VBA and VBF types, but for a comparison purpose the results here are restricted to the VBF type only.
The curves for $e^+e^- \to \nu_e \bar{\nu}_e H_5^0$ and $e^+e^- \to e^+e^- H_5^0$ are due to $WW$ fusion and $ZZ$ fusion, respectively. The difference in the magnitude of the interaction coefficient is reflected in the difference in the cross sections.
The reader should bear in mind that, as mentioned above, these two $e^+e^-$ processes also have Feynman diagrams of the VBA type (via $e^+e^- \to Z H_5^0$ and the subsequent $Z$ decay), thus for phenomenological purposes such contributions need to be taken into account.

In Fig.~\ref{fig:h50-prod}, we have also plotted the curve for $e^-e^- \to e^-e^- H_5^0$, which originally has contributions only from the Feynman diagrams of VBF type.
In fact, we can see that the size of this cross section is almost the same as $e^+e^- \to e^+e^- H_5^0$ except for the VBA type;
the tiny deviation comes from the fact that the $e^-e^-$ scattering has two diagrams where one is the crossed diagram of the other
and the interference of the two diagrams gives a tiny but positive contribution.
In both processes, once the integral luminosity of $L=\SI{10}{ab^{-1}}$ is accumulated, more than a few tens of events can be expected in the search for neutral $H_5^0$ particles with mass less than about \SI{350}{GeV}.
However, it is necessary to evaluate the background events to determine whether the search is actually possible or not.

\begin{figure}
  \centering
  \includegraphics[width=.8\textwidth]{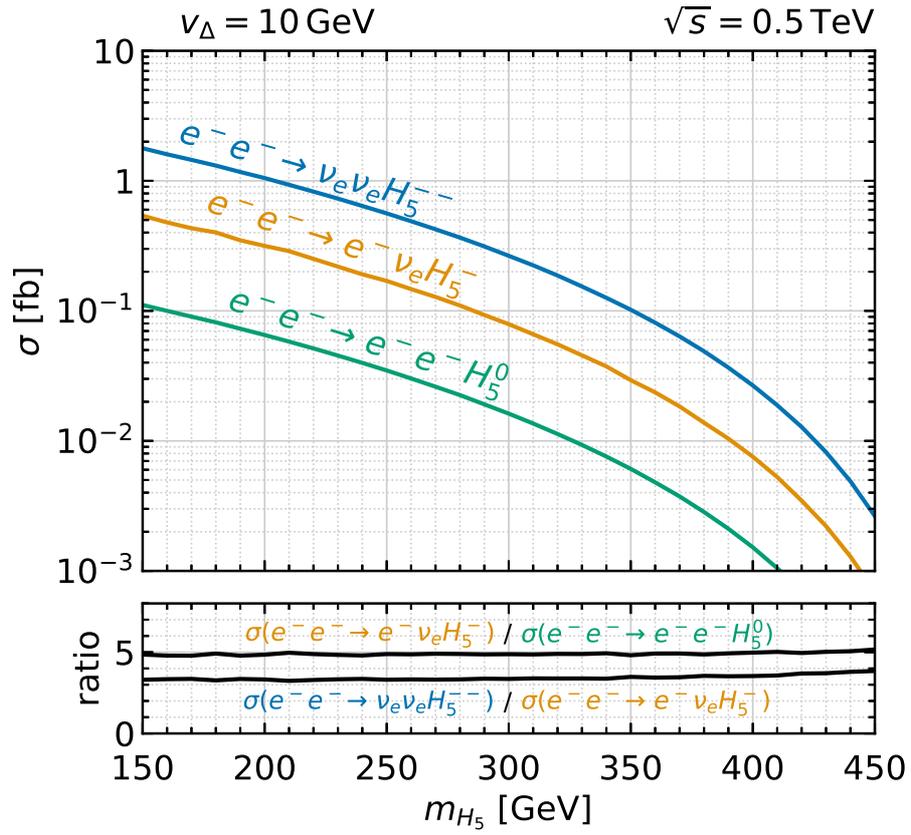}
  \caption{%
    The 5-plet Higgs boson production cross sections at $e^- e^-$ colliders.
    The ratios of the cross sections are also plotted at the bottom.
  }
  \label{fig:h5-prod}
\end{figure}

Now let us turn our attention to scattering processes in $e^- e^-$ collisions caused entirely by the VBF-type diagrams.
Fig.~\ref{fig:h5-prod} shows the production cross sections of neutral, singly- and doubly-charged $H_5$ particles at $e^-e^-$ colliders.
Since these particles are produced with a similar phase space as the VBF type, the difference in the size of the cross sections is more or less a simple dependence on the interaction constants.
The ratios of the cross sections in Fig.~\ref{fig:h5-prod} read
$\sigma(e^- e^- \to \nu_e \nu_e H_5^{--}) / \sigma(e^- e^- \to e^- \nu_e H_5^-) = \text{3.2--3.8}$ and
$\sigma(e^- e^- \to e^- \nu_e H_5^-) / \sigma(e^- e^- \to e^- e^- H_5^0) = \text{4.8--5.2}$,
and both of them gradually increase as $m_{H_5}/\sqrt{s}$ does.
These numbers are fairly close to $3$ and $9/2$, respectively,
which can be obtained by using a VBF cross-section formula in the high energy limit $\sqrt{s} \to \infty$,
see Appendix~\ref{sec:vbf-ratios}.
If such a ratio of the production cross sections is confirmed experimentally,
then it would suggest the possibility of the $H_5$ Higgs boson group in the GM model,
and therefore could be useful information for model verification.
We emphasize that such simplicity of single 5-plet Higgs boson production processes in $e^- e^-$ collisions
is because they contain only VBF-type contributions.
Single production processes in $e^+ e^-$ collisions
have more complicated mass $m_{H_5}$ dependence
due to the VBA type besides the VBF type.

\begin{figure}
  \centering
  \includegraphics[width=.8\textwidth]{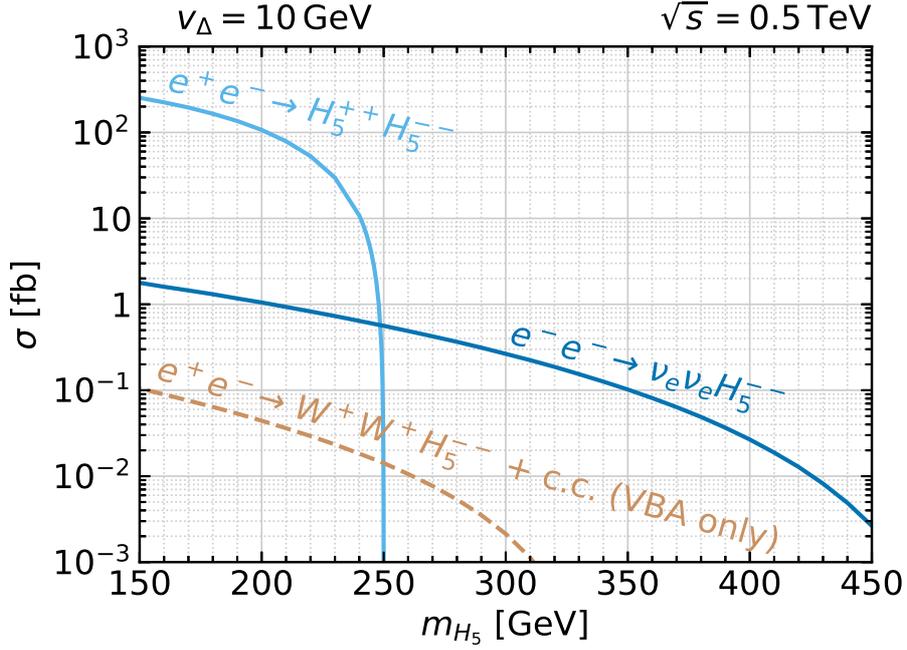}
  \caption{%
    The doubly-charged 5-plet Higgs boson production cross sections at $e^+ e^-$ and $e^- e^-$ colliders.
    The dashed curve ($e^+ e^- \to W^+ W^+ H_5^{--} + \text{c.c.}$) is
    obtained by considering only the VBA-type diagrams.
  }
  \label{fig:h5mm-prod}
\end{figure}

Since neutral or singly-charged extra Higgs bosons are predicted by many extended Higgs models,
we now turn to the doubly-charged $H_5^{--}$ particle distinctive to the GM model.
Fig.~\ref{fig:h5mm-prod} shows the calculated production cross sections of the doubly-charged $H_5^{--}$ particle in $e^+e^-$ and $e^-e^-$ collisions.
Here, in the computation for $e^+ e^- \to W^+ W^+ H_5^{--} + \text{c.c.}$,
we have omitted Feynman diagrams containing the pair production $e^+ e^- \to H_5^{++} H_5^{--}$
with the subsequent decay $H_5^{++} \to W^+ W^+$;
the calculation included only the remaining diagrams, all of which are classified as the VBA-type diagrams.
There are two targets to search for $H_5^{--}$ particles in $e^+e^-$ collisions: pair production and single production.
The cross section of the pair production is useful because of its large value due to the fact that
it is determined only by the gauge couplings~\cite{Gunion:1989ci,Chiang:2015rva}
and does not depend on $s_H$ that is now rather constrained experimentally.
This is a distinctive feature of the pair production;
all the other production cross sections plotted in
Figs.~\ref{fig:h50-prod}--\ref{fig:h5mm-prod}
are suppressed by $s_H^2$.
However, the accessible mass value in the pair production is limited to $\sqrt{s}/2$.
On the other hand, since the single production of $H_5^{--}$ in $e^+e^-$ collisions involves a $W^+W^+$ pair, the upper limit of the search mass range is $\sqrt{s}-2 m_W$;
there is no significant extension in comparison with the pair production for a moderate value of $\sqrt{s}$.
By contrast, in $e^-e^-$ collisions, there is a process $e^-e^- \to \nu_e \nu_e H_5^{--}$ that has a wider search mass range.
Specifically, in this process, the search range is extended to $m_{H_5}\sim \sqrt{s}$ in principle.
In fact, we can see from Fig.~\ref{fig:h5mm-prod} that
about 100 events are expected for $L=\SI{10}{ab^{-1}}$ of integrated luminosity
when $m_{H_5}\sim \SI{0.4}{TeV}$ with $\sqrt{s}=\SI{0.5}{TeV}$.
Note that the cross section for $e^- e^- \to \nu_e \nu_e H_5^{--}$
in Fig.~\ref{fig:h5mm-prod} is more than 6 times larger than that for
$e^+ e^- \to \nu_e \bar{\nu}_e H_5^0$ (VBF only) in Fig.~\ref{fig:h50-prod};
the difference between them comes entirely from that
in the Higgs couplings to two $W$ bosons,
except in the interference term in the former.

\begin{figure}
  \centering
  \includegraphics[width=.8\textwidth]{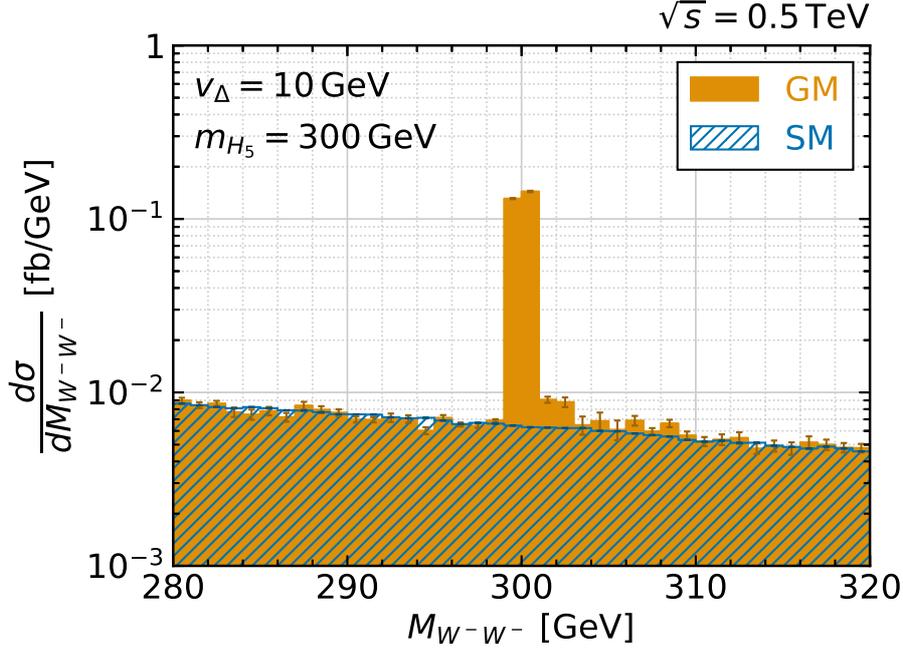}
  \caption{%
    The $W^- W^-$ invariant mass distribution for $e^- e^- \to \nu_e \nu_e W^- W^-$.
    Each error bar indicates the statistical error of the Monte Carlo integration for each bin.
  }
  \label{fig:ww-dist}
\end{figure}

If the $H_5^{--}$ particle decays mainly into a $W^-W^-$ pair,
then we expect to observe a resonance peak at $M_{W^- W^-}\sim m_{H_5}$ in the invariant mass $M_{W^- W^-}$ distribution of $W^-$ pairs in the $e^- e^-\to \nu_e \nu_e W^- W^-$ process
(see the resonant diagram Fig.~\ref{fig:ee-diag}d).
In Fig.~\ref{fig:ww-dist}, we plot
the $M_{W^- W^-}$ distribution with $m_{H_5} = \SI{300}{GeV}$ for that decay branching ratio of $100$\%,%
\footnote{%
  $\text{BR}(H_5^{\pm\pm} \to W^\pm W^\pm) \approx 1$ is usually assumed in analyses for the LHC~\cite{Zaro:2015ika},
  which is true if $\Delta m \equiv m_{H_5}-m_{H_3}$ is smaller than $m_W$.
  For $\Delta m > m_W$, the decay branching ratio of $H_5^{--} \to W^- H_3^-$ is no longer zero.
  The details of the signal in this case are currently under analysis.
  In general, the result depends on $m_{H_3}$ and the mixing angle $\alpha$
  (if a subsequent $H_3^- \to h W^-$ decay is considered),
  and is beyond the scope of this paper, thus will be presented elsewhere.
}
along with the contribution from the SM\@.
Admittedly, a similar plot was obtained in Ref.~\cite{Barger:1994wa}, but here we update the plot considering the current experimental constraints.

\begin{figure}
  \centering
  \includegraphics[width=.8\textwidth]{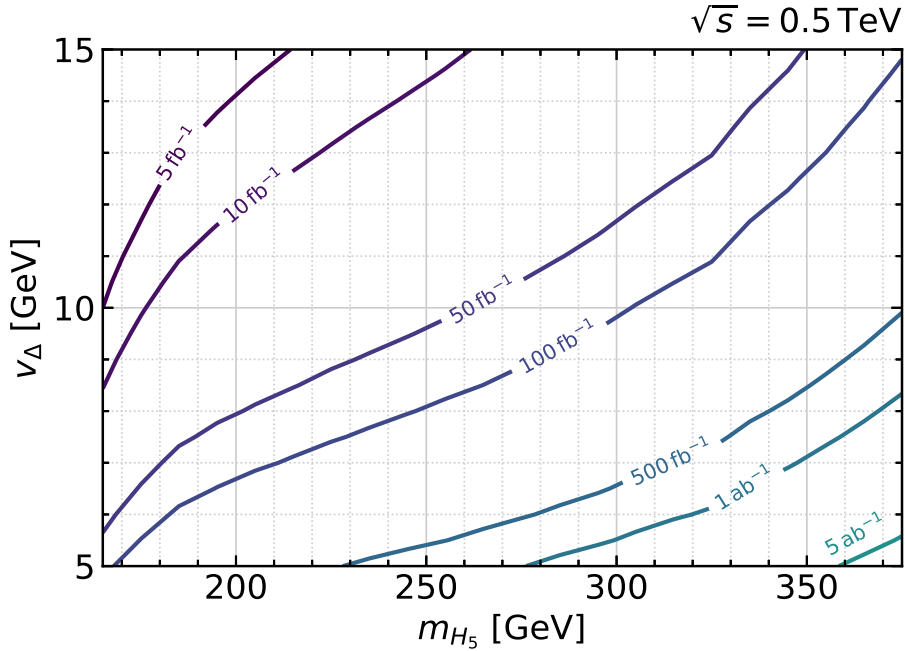}
  \caption{%
    The integrated luminosity required for a 5-$\sigma$ discovery
    by the $W^- W^-$ invariant mass distribution
    for $e^- e^- \to \nu_e \nu_e W^- W^-$.
    A common invariant mass window cut
    $m_{H_5} - \SI{20}{GeV} \le M_{W^-W^-} \le m_{H_5} + \SI{20}{GeV}$
    is used for every $m_{H_5}$
    and $\text{BR}(H_5^{--} \to W^- W^-) = 1$ is assumed.
  }
  \label{fig:ww-lumi}
\end{figure}

Note that, unlike hadron colliders such as the LHC and LHeC~\cite{Sun:2017mue}, the invariant mass $M_{W^- W^-}$ can be cleanly reconstructed using the $4$-jets from $W^- \to q \bar{q}'$ decays in the $e^- e^-$ collider.
The total cross section of the $W^-$ pair with missing energy in the SM is about \SI{2.5}{fb}, whereas the total cross section of the signal in the GM model is about \SI{2.8}{fb}.
From this result, it is clear that if we simply consider only the statistical error in the value of the total cross section and the number of events,%
\footnote{%
  For simplicity, we use $Z = S / \sqrt{B}$ as an estimator
  to compute the signal significance $Z$,
  where $S$ and $B$ are the numbers of signal and background events, respectively.
}
an integral luminosity of about \SI{1}{ab^{-1}} is required to obtain an event excess of \SI{5}{\sigma}, while about \SI{100}{fb^{-1}} is sufficient when the analysis is limited to the region of Fig.~\ref{fig:ww-dist}.
We also show a contour plot in the $m_{H_5}$-$v_\Delta$ plane, in Fig.~\ref{fig:ww-lumi},
for the integrated luminosity required for a 5-$\sigma$ discovery
via a signal in the $W^- W^-$ invariant mass distribution of $e^- e^- \to \nu_e \nu_e W^- W^-$.
For this plot, we have used a common invariant mass window cut
$m_{H_5} - \SI{20}{GeV} \le M_{W^-W^-} \le m_{H_5} + \SI{20}{GeV}$
for every $m_{H_5}$ and assumed all $H_5^{--}$ decay into $W^- W^-$ pairs.

Considering how to search for the 5-plet Higgs bosons in the GM model
using VBF processes at future linear colliders,
especially via the $WW$ fusion processes
because of their relatively large sizes of the cross sections,
one immediately sees the following fact:
due to the electric charges,
$e^+ e^-$ colliders can produce the $H_5^0$ bosons in this channel,
while $e^- e^-$ colliders produce the $H_5^{--}$ bosons.
Therefore, a search strategy can be developed as follows.
With a future linear collider running under an $e^+ e^-$ mode,
we first target the $H_5^0$ boson by using
$e^+ e^- \to H_5^0 + \text{invisible}$
with a subsequent $H_5^0$ decay into a $W^+ W^-$ or $Z Z$ pair,
assuming $\text{BR}(H_5^0 \to VV) \approx 1$.
If found, then this will be certainly an outstanding triumph,
but countless models are predicting such a new neutral scalar particle.
Then, suppose the linear collider has an $e^- e^-$ mode as an option.
We next run the collider under the $e^- e^-$ mode
in order to target the $H_5^{--}$ boson by using
$e^- e^- \to H_5^{--} + \text{invisible}$
with a subsequent $H_5^{--}$ decay into a $W^- W^-$ pair.
If nature adopts the GM model
and the discovered extra neutral Higgs boson is indeed $H_5^0$,
then we will find a resonant peak by $H_5^{--}$ at the same mass,
strongly suggesting that the model may be correct.
This will hopefully be further confirmed by the search for $H_5^-$ in $e^- e^-$ collisions.

\section{Summary}
\label{sec:summary}

We have discussed our extension of the \GRACE{} system to perform calculations
in the Georgi--Machacek (GM) model.
After a systematic check of the implemented set of model files
against the known results of the $1 \to 2$ and $2 \to 2$ processes,
we have analyzed the fermiophobic custodial 5-plet Higgs boson production processes
in the GM model at $e^+ e^-$ and $e^- e^-$ colliders with
the center-of-mass energy of $\sqrt{s} = \SI{0.5}{TeV}$.
The mass of $m_{H_5}$ and the triplet VEV $v_\Delta$ have been chosen as
typical values that are not currently ruled out by the experimental data.

The neutral 5-plet Higgs boson $H_5^0$ can be produced in several processes
both at $e^+ e^-$ and $e^- e^-$ colliders,
provided $m_{H_5} < \sqrt{s}$.
As shown in Fig.~\ref{fig:h50-prod},
the largest cross section is obtained for $e^+ e^- \to Z H_5^0$.
The other processes have smaller cross sections, however, even the $ZZ$ fusion processes
possibly lead to enough events to be detected
with the integrated luminosity expected at future colliders.
We have also pointed out that at $e^- e^-$ colliders
the sizes of the neutral, singly- and doubly-charged 5-plet Higgs boson production cross sections
via vector boson fusions
have a relatively simple dependence of coupling constants.
The ratios of the cross sections are fairly close to
those estimated in the high energy limit even at $\sqrt{s} = \SI{0.5}{TeV}$.

Since the doubly-charged Higgs boson $H_5^{\pm\pm}$ is one of the characteristic particles
in the GM model, we have performed computations for its production,
for which $e^- e^-$ colliders are rather suitable
from the viewpoint of the quantum number of the initial state.
While the $H_5^{++}$ $H_5^{--}$ pair production at $e^+ e^-$ colliders can be used
for searches with the mass of $m_{H_5}$ only up to $\sqrt{s}/2$,
the single $H_5^{--}$ production at $e^- e^-$ colliders can reach heavier mass regions.
We have also investigated the resonant $M_{W^-W^-}$ distribution in $e^- e^- \to \nu_e \nu_e W^- W^-$
with an assumption that the decay branching ratio for $H_5^{--} \to W^- W^-$ is 100\%.

In this paper, we have focused on the 5-plet Higgs boson production,
but one can also consider 3-plet and singlet Higgs boson production at $e^- e^-$ colliders.
We leave it for future work.

\section*{Acknowledgment}

We are grateful to Takuto Nagura for discussions and
for his participation in this study at the early stage.
The work by T.U. is in part supported by JSPS KAKENHI Grant Numbers 19K03831 and 21K03583.

\appendix

\section{Formulae}
\label{sec:formulae}

In the following formulae we use variables $x_A \equiv {m_A^2}/{M_H^2}$, $z_A \equiv {m_A^2}/{s}$ and the function $\lambda(x,y)\equiv 1+x^2+y^2-2x-2y-2xy$. $N_c$ is the color factor of the final state fermions.

\begin{itemize}

\item
Decay widths for $H \to V V'$, $H \to H' V$ and $H \to f\bar{f}'$
\begin{equation}
 \Gamma(H \to VV')
  = {\frac{g_{HVV'}^2}{8 \pi M_H}} \lambda^\frac{1}{2}\left(x_V, x_{V'}\right)
\left[1+{\frac{\left(1 - x_{V}-x_{V'}\right)^2}{8 x_V x_{V'}}} \right],
\end{equation}
\begin{equation}
 \Gamma(H \to H'V)
  = {\frac{g_{HH'V}^2}{16 \pi}} M_H \lambda^\frac{1}{2}\left(x_{H'}, x_{V}\right)
\left[{\frac{\left( 1-x_{H'}+x_V \right)^2}{x_V}}-4 \right],
\end{equation}
\begin{equation}
  \begin{split}
    \Gamma(H \to f\bar{f}')
    &= {\frac{N_c}{16 \pi}} M_H \lambda^\frac{1}{2}\left(x_{f}, x_{f'}\right)
    \\ & \qquad
    \times\biggl[ \left( \left| g_{Hff'}^L\right|^2+ \left| g_{Hff'}^R \right|^2 \right)
    \left( 1-x_f - x_{f'}\right)
    \\ & \qquad\qquad
    - 4 \Re\left(g_{Hff'}^L g_{Hff'}^{R*} \right) \sqrt{x_f x_{f'}} \biggr].
  \end{split}
\end{equation}

\item
Differential cross sections for $f\bar{f}' \to HH'$
	\begin{itemize}

	\item
$(f, \bar{f}')=(e^-_{L,R}, e^+)$ and $(H, H') = (H_5^{++},H_5^{--}), (H_5^{+},H_5^{-}),(H_3^{+},H_3^{-})$
	\begin{equation}
 {\frac{d\sigma_{L,R}}{d\cos\theta}} = {\frac{\lambda^{\frac{3}{2}}\left(z_H, z_{H'}\right)}{64 \pi s}}
\left( g_{HH'\gamma}g_{ff'\gamma}+{\frac{g_{HH'Z}g_{ff'Z}^{L,R}}{1-z_Z}} \right)^2
\sin^2\theta.
	\end{equation}

	\item
$(f, \bar{f}')=({\nu}_{eL}, e^+)$ and $(H, H') = (H_5^{++},H_5^{-}), (H_5^{+},H_5^{0}),(H_3^{+},H_3^{0})$
	\begin{equation}
 {\frac{d\sigma_{L}}{d\cos\theta}} = {\frac{\lambda^{\frac{3}{2}}\left(z_H, z_{H'}\right)}{32 \pi s}} \left({\frac{g_{HH'W}g_{ff'W}^{L}}{1-z_W}} \right)^2
 \sin^2\theta.
	\end{equation}

	\end{itemize}

\item
Differential cross sections for $f\bar{f}' \to Vh^0$
  \begin{itemize}

  \item $(f, \bar{f}')=(e^-_{L,R}, e^+)$ and $V = Z$
	\begin{equation}
    \begin{split}
      {\frac{d\sigma_{L,R}}{d\cos\theta}}
      &=
      {\frac{\lambda^{\frac{1}{2}}(z_V, z_{h})}{64 \pi s^2}}
      \left(\frac{g_{hVV}g_{ff'V}^{L,R}}{1-z_V}\right)^2
      \left[ 2+{\frac{\lambda(z_V,z_h)}{4 z_V}}\sin^2\theta \right].
    \end{split}
    \label{eq:A6}
	\end{equation}

  \item Eq.~\eqref{eq:A6} can be applied also for $(f, \bar{f}')=({\nu}_{eL}, e^+)$ and $V = W^+$,
        but then an extra factor 2 must be multiplied.

  \end{itemize}

	\item
Coupling constants
  \allowdisplaybreaks[4]
  \begin{align}
    g_{H_5^{\pm\pm}W^\mp W^\mp}&={\frac{g^2}{\sqrt{2}}} s_H v,
    \\
    g_{H_5^{\pm}W^\mp Z}&=\mp {\frac{g g_Z}{2 }} s_H v,
    \\
    g_{H_5^{0}W^+ W^-}&=-{\frac{g^2}{2\sqrt{3}}} s_H v,
    \\
    g_{H_5^{0}Z Z}&={\frac{g_Z^2}{\sqrt{3}}} s_H v,
    \\
    g_{H_1^{0}W^+ W^-}&=-{\frac{g^2}{6}} \left(3 s_\alpha c_H -2\sqrt{6} c_\alpha s_H\right) v,
    \\
    g_{H_1^{0}Z Z}&=-{\frac{g_Z^2}{6}} \left(3 s_\alpha c_H -2\sqrt{6} c_\alpha s_H\right) v,
    \\
    g_{h^{0}W^+ W^-}&={\frac{g^2}{6}} \left(3 c_\alpha c_H +2\sqrt{6} s_\alpha s_H\right) v,
    \\
    g_{h^{0}Z Z}&={\frac{g_Z^2}{6}} \left(3 c_\alpha c_H +2\sqrt{6} s_\alpha s_H\right) v,
    \\
    g_{H_5^{++}H_5^{--} \gamma} &= 2 e,
    \\
    g_{H_5^{+}H_5^{-} \gamma} &= -e,
    \\
    g_{H_3^{+}H_3^{-} \gamma} &= -e,
    \\
    g_{H_5^{++}H_5^{--} Z} &= g_Z \left(1-2 s_W^2 \right),
    \\
    g_{H_5^{+}H_5^{-} Z} &= -{\frac{g_Z}{2}} \left(1-2 s_W^2 \right),
    \\
    g_{H_3^{+}H_3^{-} Z} &= -{\frac{g_Z}{2}} \left(1-2 s_W^2 \right),
    \\
    g_{H_5^{\pm}H_3^{\mp} Z} &= \pm{\frac{g_Z}{2}} c_H,
    \\
    g_{H_5^{0}H_3^{0} Z} &= i{\frac{g_Z}{\sqrt{3}}} c_H,
    \\
    g_{H_3^{0}H_1^{0} Z} &= -i{\frac{g_Z}{6}} \left(3 s_\alpha s_H +2\sqrt{6} c_\alpha c_H\right),
    \\
    g_{H_5^{\pm\pm}H_5^{\mp} W^{\mp}} &= -{\frac{g}{\sqrt{2}}},
    \\
    g_{H_5^{\pm}H_5^{0} W^{\mp}} &= {\frac{\sqrt{3}}{2}}g,
    \\
    g_{H_5^{\pm\pm}H_3^{\mp} W^{\mp}} &= -{\frac{g}{\sqrt{2}}}c_H,
    \\
    g_{H_5^{\pm}H_3^{0} W^{\mp}} &= \mp i{\frac{g}{2}}c_H,
    \\
    g_{H_3^{\pm}H_5^{0} W^{\mp}} &= -{\frac{g}{2\sqrt{3}}}c_H,
    \\
    g_{H_3^{\pm}H_3^{0} W^{\mp}} &= \mp i{\frac{g}{2}},
    \\
    g_{H_3^{\pm}H_1^{0} W^{\mp}} &= {\frac{g}{6}} \left(3 s_\alpha s_H +2\sqrt{6} c_\alpha c_H\right),
    \\
    g_{H_3^{\pm}h^{0} W^{\mp}} &= -{\frac{g}{6}} \left(3 c_\alpha s_H -2\sqrt{6} s_\alpha c_H\right),
    \\
    g_{H_3^{0}h^{0} Z} &= i{\frac{g_Z}{6}} \left(3 c_\alpha s_H -2\sqrt{6} s_\alpha c_H\right),
    \\
    g_{H_3^{+,-}ud}^L &= -\sqrt{2}V_{ud}{\frac{m_{u,d}}{v}} t_H,
    \\
    g_{H_3^{+,-}ud}^R &= \sqrt{2}V_{ud}{\frac{m_{d,u}}{v}} t_H,
    \\
    g_{H_3^{+}e\nu}^R &= \sqrt{2}{\frac{m_{e}}{v}} t_H,
    \\
    g_{H_3^{-}e\nu}^L &= \sqrt{2}{\frac{m_{e}}{v}} t_H,
    \\
    g_{H_3^{0}uu}^L &= i{\frac{m_u}{v}} t_H,
    \\
    g_{H_3^{0}uu}^R &= -i{\frac{m_u}{v}} t_H,
    \\
    g_{H_3^{0}dd}^L &= -i{\frac{m_d}{v}} t_H,
    \\
    g_{H_3^{0}dd}^R &= i{\frac{m_d}{v}} t_H,
    \\
    g_{H_3^{0}ee}^L &= -i{\frac{m_e}{v}} t_H,
    \\
    g_{H_3^{0}ee}^R &= i{\frac{m_e}{v}} t_H,
    \\
    g_{H_1^{0}uu}^{L,R} &= {\frac{m_u}{v}} \frac{s_\alpha}{c_H},
    \\
    g_{H_1^{0}dd}^{L,R} &= {\frac{m_d}{v}} \frac{s_\alpha}{c_H},
    \\
    g_{H_1^{0}ee}^{L,R} &= {\frac{m_e}{v}} \frac{s_\alpha}{c_H},
    \\
    g_{h^{0}uu}^{L,R} &= -{\frac{m_u}{v}} \frac{c_\alpha}{c_H} ,
    \\
    g_{h^{0}dd}^{L,R} &= -{\frac{m_d}{v}} \frac{c_\alpha}{c_H},
    \\
    g_{h^{0}ee}^{L,R} &= -{\frac{m_e}{v}} \frac{c_\alpha}{c_H}.
  \end{align}
  \allowdisplaybreaks[0]
  Here, we have used additional shorthand notations
  $g_Z = g / \cos\theta_W$ and
  $s_W = \sin\theta_W$.

\end{itemize}

\section{Ratios of VBF total cross sections}
\label{sec:vbf-ratios}

The typical Feynman diagram contribution for VBF processes
can be written as, in the Feynman gauge,
\begin{equation}
  \mathcal{M}_A
  =
  i g_{V_1V_2H}
  \frac{
    \bigl[ \bar{u}(p_3) \gamma^\mu (g_{R_1} \mathcal{P}_R + g_{L_1} \mathcal{P}_L) u(p_1) \bigr]
    \bigl[ \bar{u}(p_4) \gamma_\mu (g_{R_2} \mathcal{P}_R + g_{L_2} \mathcal{P}_L) u(p_2) \bigr]
  }{
    \bigl[ (p_1-p_3)^2 - m_{V_1}^2 \bigr]
    \bigl[ (p_2-p_4)^2 - m_{V_2}^2 \bigr]
  }.
\end{equation}
Here, an incoming fermion with the momentum $p_1$ splits into a fermion with the momentum $p_3$
and the gauge boson of mass $m_{V_1}$ via the interaction specified by $g_{R_1}$ and $g_{L_1}$.
The same for the other side; namely, the other incoming fermion with the momentum $p_2$
splits into a fermion with the momentum $p_4$ and the gauge boson of mass $m_{V_2}$
via the interaction specified by $g_{R_2}$ and $g_{L_2}$.
The Higgs boson of mass $m_H$ is produced by the fusion of the two gauge bosons with the coupling $g_{V_1V_2H}$.
All fermion masses are ignored.

For $e^- e^-$ collisions, one should also consider a crossed diagram, obtained by the exchange of
the incoming electrons, $\mathcal{M}_B = - (\mathcal{M}_A \text{ with $p_1 \leftrightarrow p_2$})$.
One can argue that the interference term between $\mathcal{M}_A$ and $\mathcal{M}_B$ is negligible~\cite{Dicus:1985zg}.%
\footnote{%
  In fact, this was numerically observed in Ref.~\cite{Hikasa:1985ee} for
  $e^- e^- \to e^- e^- Z Z \to e^- e^- h$ in the SM\@.
  See also the almost negligible deviation between the two curves for the $ZZ$ fusion
  at $e^+e^-$ and $e^-e^-$ colliders in Fig.~\ref{fig:h50-prod} and its explanation in the text.
}
If one is interested only in the total cross section,
$\lvert \mathcal{M}_B \rvert^2$ gives the same contribution as $\lvert \mathcal{M}_A \rvert^2$,
thus a factor 2 is multiplied.

As a crude approximation in the high energy limit ($\sqrt{s} \gg m_H, m_{V_1}, m_{V_2}$),
one can drive a Weizs\"acker--Williams-type expression for the total cross section~%
\cite{Kane:1984bb,Chanowitz:1985hj,Dawson:1984gx,Altarelli:1987ue}.%
\footnote{%
  The leading logarithmic term was obtained in Ref.~\cite{Cahn:1983ip},
  which is indeed sufficient to obtain the ratios in Eqs.~\eqref{eq:ratio_WW/ZW} and~\eqref{eq:ratio_ZW/ZZ}.
  The approximation can be in principle improved by eliminating kinematical oversimplifications,
  see, e.g., Ref.~\cite{Johnson:1987tj}.
}
The result for $e^- e^-$ collisions is
\begin{equation}
  \sigma \simeq \frac{\mathcal{S} g_{V_1V_2H}^2 C}{2 (4\pi)^3 m_{V_1}^2 m_{V_2}^2}
  \left[ (1 + z_H) \ln\left(\frac{1}{z_H}\right) - 2 + 2 z_H \right],
\end{equation}
where $\mathcal{S}$ represents the statistical factor for the outgoing fermions, $z_H = m_H^2 / s$ and
\begin{equation}
  C = (g_{R_1}^2 + g_{L_1}^2) (g_{R_2}^2 + g_{L_2}^2).
\end{equation}
Although this formula overestimates the total cross section by a factor $\gtrapprox 2$
for $\sqrt{s} = \SI{0.5}{TeV}$,
one may expect that such deviations cancel for the most part when ratios are taken.
With this formula, the ratios of the VBF cross sections at $e^- e^-$ colliders
presented in Sec.~\ref{sec:result} are estimated as
\begin{align}
  \frac{\sigma(e^- e^- \to \nu_e \nu_e H_5^{--})}{\sigma(e^- e^- \to e^- \nu_e H_5^{-})}
    & \simeq
  2
  \left( \frac{\cos^2\theta_W}{1 - 4 \sin^2\theta_W + 8 \sin^4\theta_W} \right)
  \approx
  3,
  \label{eq:ratio_WW/ZW}
  \\
  \frac{\sigma(e^- e^- \to e^- \nu_e H_5^{-})}{\sigma(e^- e^- \to e^- e^- H_5^0)}
    & \simeq
  3
  \left( \frac{\cos^2\theta_W}{1 - 4 \sin^2\theta_W + 8 \sin^4\theta_W} \right)
  \approx
  \frac{9}{2}.
  \label{eq:ratio_ZW/ZZ}
\end{align}
Note that the couplings between $H_5$ and two vector bosons arise from
the kinetic term of the isospin triplet,
picking up the triplet VEV\@.
Thus, all of them are proportional to $s_H v$,
resulting in the above ratios consisting of group theoretical factors and the weak mixing angle.
In general, if one considers other ratios, e.g., of production cross sections of $H_5$ and $H_1$,
then the other mixing angle $\alpha$ appears as well as $\theta_H$.

\bibliographystyle{JHEP-mod}
\bibliography{references}

\end{document}